\documentclass[%
 reprint,
 twocolumn,
 superscriptaddress,
 amsmath,amssymb,
 aps,
]{revtex4-2}

\usepackage{graphicx}
\usepackage{dcolumn}
\usepackage{bm}
\usepackage{color} 
\usepackage{float}
\makeatletter
\let\newfloat\newfloat@ltx
\makeatother
\usepackage{algorithm} 
\usepackage{algpseudocode} 
\usepackage{hyperref}

\begin{document}


\title{Deep learning based parameter search for an agent based social network model}%

\author{Yohsuke Murase}
\affiliation{RIKEN Center for Computational Science, Kobe, Japan}
\author{Hang-Hyun Jo}
\email{h2jo@catholic.ac.kr}
\affiliation{Department of Physics, The Catholic University of Korea, Bucheon, Republic of Korea}%
\author{J\'anos T\"or\"ok}
\affiliation{Department of Theoretical Physics, Budapest University of Technology and Economics, Budapest, Hungary}
\affiliation{Department of Network and Data Science, Central European University, Vienna, Austria}
\affiliation{MTA-BME Morphodynamics Research Group, Budapest University of Technology and Economics, Budapest, Hungary}
\author{J\'anos Kert\'esz}
\affiliation{Department of Network and Data Science, Central European University, Vienna, Austria}
\affiliation{Department of Computer Science, Aalto University, Espoo, Finland}
\author{Kimmo Kaski}
\affiliation{Department of Computer Science, Aalto University, Espoo, Finland}
\affiliation{The Alan Turing Institute, British Library, London, UK}

\date{\today}

\begin{abstract}
Interactions between humans give rise to complex social networks that are characterized by heterogeneous degree distribution, weight-topology relation, overlapping community structure, and dynamics of links. Understanding such networks is a primary goal of science due to serving as the scaffold for many emergent social phenomena from disease spreading to political movements. An appropriate tool for studying  them is agent-based modeling, in which nodes, representing persons, make decisions about creating and deleting links, thus yielding various macroscopic behavioral patterns. Here we focus on studying a generalization of the weighted social network model, being one of the most fundamental agent-based models for describing the formation of social ties and social networks. This Generalized Weighted Social Network (GWSN) model incorporates triadic closure, homophilic interactions, and various link termination mechanisms, which have been studied separately in the previous works. Accordingly, the GWSN model has an increased number of input parameters and the model behavior gets excessively complex, making it challenging to clarify the model behavior. We have executed massive simulations with a supercomputer and using the results as the training data for deep neural networks to conduct regression analysis for predicting the properties of the generated networks from the input parameters. The obtained regression model was also used for global sensitivity analysis to identify which parameters are influential or insignificant. We believe that this methodology is applicable for a large class of complex network models, thus opening the way for more realistic quantitative agent-based modeling.
\end{abstract}

\maketitle

\section{Introduction}

When analyzing the structural patterns of real social networks, the synthetic model-generated networks have served as references for comparison and enhanced insight into their properties~\cite{kertesz2021modeling}. Here we classify the models of social networks into two categories: static models and dynamic models. The static models constitute a family of models, in which the network links are randomly generated with certain constraints. The most fundamental one is the Erd\"os-R\'enyi model~\cite{Erdos1960Evolution}, which generates a random network under the constraint on the average degree. Other examples include the configuration model, the exponential random graph model, and the stochastic block models~\cite{Newman2018Networks, Barabasi2016Network, Menczer2020First}. Often these models enable us to compute their properties analytically. After inferring the model parameters for a real social network of interest, these models serve as useful null models. One can then judge, for instance, whether an observed quantity is significantly different from the expected null model value or not, thus serving as hypothesis testing. 

On the other hand, dynamic models are models in which the network evolves as a function of time. One of the major objectives of the dynamic models is to find the mechanisms that lead to certain structures observed in empirical networks. Here the models are defined by the rules on how nodes and links are created or deleted, in order to incorporate the perceived mechanisms of the evolutionary processes of social networks. However, in the light of vast complexities of social dynamics, these mechanisms should, at best, be considered as plausible ones. This class includes Barab\'asi-Albert scale-free network model and its generalizations~\cite{Barabasi2016Network} as well as Kumpula et al.'s weighted social network model~\cite{Kumpula2007Emergence}. The latter will be in the focus of the present paper. These models allow us to get insight into how and why the observed networks have been generated and, more importantly, to predict the possible evolution of real networks.

While the static models are often analytically solvable, analytical tractability of dynamic models is limited to basic cases. These models are usually designed to be as simple as possible, in order to identify the most important mechanisms, but they are not suitable for quantitative comparison and prediction. When models are extended to incorporate aspects of reality, understanding their behavior becomes a formidable task since the parameter space is high-dimensional and non-trivial relationships between the parameters may occur. This makes the choice of appropriate parameters very difficult.

In this paper, we are going to overcome this difficulty by top scale computing approach and by the development of a meta-model in order to investigate the parameter space of an agent-based model of social networks. To achieve this a massive number of simulations will be performed, the results of which are then used as training data for a neural network model for inferring or analysing the properties of the generated networks. Such a regression model is called a meta-model or a surrogate model as it is a model of a simulation model~\cite{Ghiasi2018comparative, Zhao2010comparative, Wang2007Review}. Meta-models are developed as approximations of the expensive simulation process in order to improve the overall computation efficiency and they are found to be a valuable tool to support activities in modern engineering design, especially design optimization. Once a good meta-model is obtained, it is useful for various purposes including parameter tuning, understanding of the model behavior, and sensitivity analysis. The meta-model is effective especially when the simulations are computationally demanding, which is the case for agent-based models with many parameters. To the best of our knowledge, this study is the first attempt to apply the meta-modeling approach to an agent-based model of social networks. Here we will also carry out a sensitivity analysis to demonstrate that this approach is useful to distinguish between influential and negligible parameters.

This paper is organized as follows. In Section~\ref{sec:gwsn}, we first introduce Kumpula et al.'s weighted social network model and its various extensions, then we formulate a generalized model to incorporate most of the previous extensions. In Section~\ref{sec:analysis}, we conduct the regression analysis and the sensitivity analysis using neural networks. The last Section is devoted to the summary and discussion.

\begin{figure*}[!t]
\begin{center}
\includegraphics[width=0.9\textwidth]{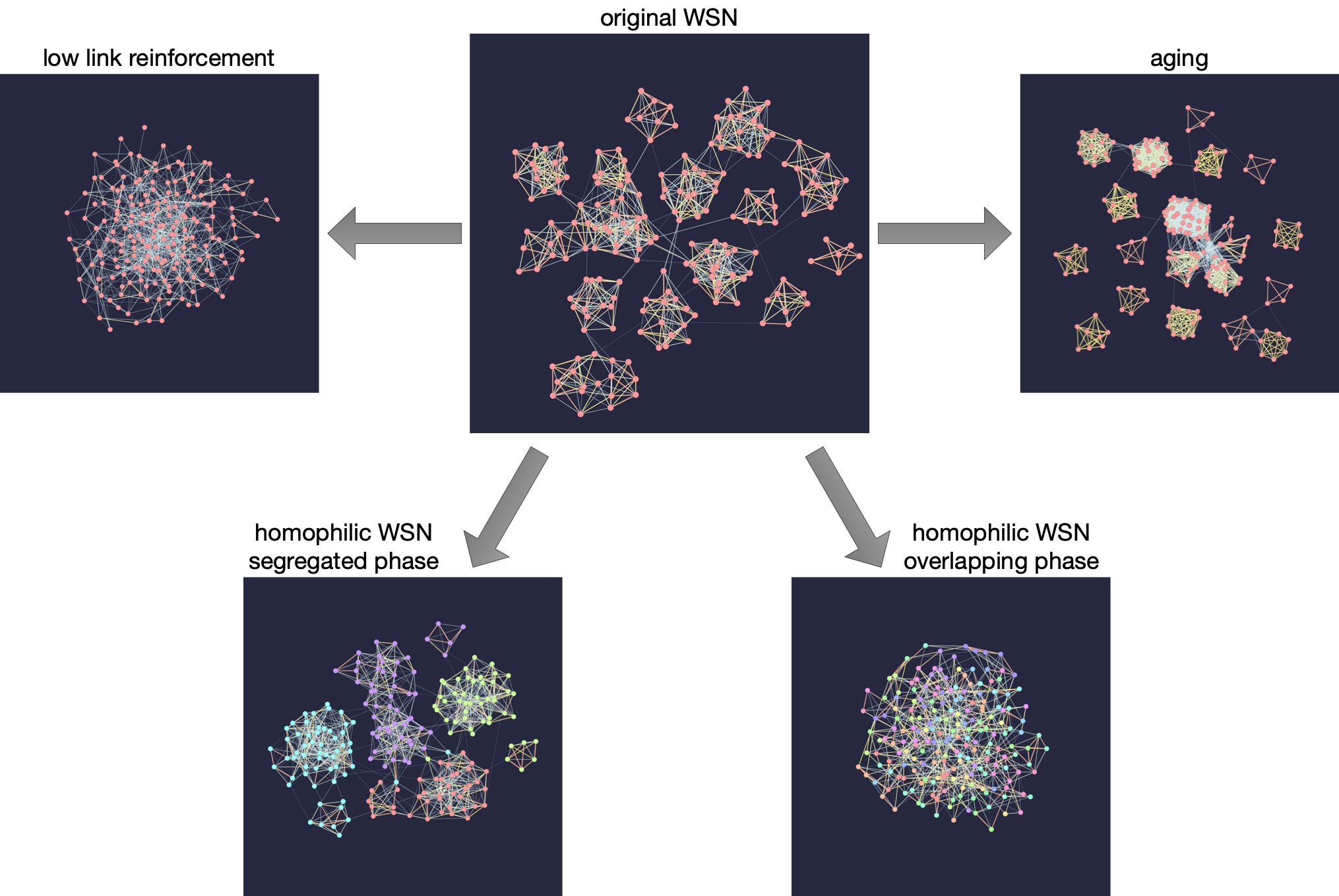}
\end{center}
\caption{Snapshots of the networks generated by the GWSN model for various combinations of parameter values in the model. The top center panel shows the network generated by the original WSN model showing the Granovetterian community structure when the links are reinforced sufficiently. The Granovetterian community structure is absent when the link reinforcement is not high enough, as shown in the top left panel. The top right panel presents the network with higher clustering than the original WSN model when only the link aging mechanism is applied for the link termination. The bottom panels show the effect of homophilic interaction on the generated networks; depending on the parameter values, it shows the structural transition from the segregated, non-overlapping communities (bottom left) to the overlapping communities (bottom right).}
\label{fig:snapshot}
\end{figure*}

\section{Generalized Weighted Social Network Model}\label{sec:gwsn}

\subsection{Background}

The original weighted social network (WSN) model was introduced in Ref.~\cite{Kumpula2007Emergence}, which is a dynamic, undirected, weighted network model leading to a stationary state. The model starts with an empty network of $N$ nodes. Links are created, updated, and eliminated using three essential mechanisms, which are called a global attachment (GA), a local attachment (LA), and a node deletion (ND). The GA mechanism represents the focal closure in sociology~\cite{Kossinets2006Empirical}, meaning that ties between people are formed due to the shared activities. In the WSN model it is implemented by creating links between randomly chosen nodes. The LA mechanism represents the cyclic or triadic closure~\cite{Kossinets2006Empirical}, implying that friends of friends get to know each other. The created new links by GA or LA mechanisms are assigned link weights of the same positive value. Whenever the LA mechanism is applied, weights of involved links increase by some fixed non-negative value, implying the link reinforcement (note that the triadic closure is not always successful in the LA mechanism, which will be discussed later). Finally, the ND mechanism represents a turnover of nodes, which can be implemented by removing all the links incident to a randomly chosen node. Without the ND mechanism, the model would end up with a complete graph, while with it a stationary state sets in for long times. 

The WSN model turns out to generate several stylized facts of social networks~\cite{Jo2018Stylized}, among which the Granovetterian community structure is the most important one as it connects the global structure of the network with local features~\cite{Granovetter1973Strength}. The Granovetterian community structure means that communities of nodes that are tightly and strongly connected to each other are connected via weak ties, as shown in Fig.~\ref{fig:snapshot}(top center); nodes connected by strong links within communities have many common neighbors while the opposite is true for weak links. If one of the above mechanisms is missing from the model, the model is not able to reproduce the Granovetterian community structure, indicating that the model contains the minimal set of the essential ingredients for this structural feature. Among several parameters in the model, link reinforcement is the most crucial one to get the desired community structure (see Fig.~\ref{fig:snapshot}(top left)).

Let us briefly sketch how the Granovetterian community structure emerges in the model: As the nodes are initially isolated, they make connections with other nodes by means of the GA mechanism. By the LA mechanism, created links are gradually strengthened and occasionally new links are also formed by the triadic closure. Such clusters formed around triangles are seeds for communities. These seeds develop as their sizes increase by the GA mechanism, and the links inside the communities get more dense due to the triadic closure and their weights become larger by the link reinforcement, thus leading to the Granovetterian community structure. By the ND mechanism, all the links adjacent to a randomly chosen node are removed in order to reach the nontrivial stationary state of network dynamics.

Although the WSN model has remarkably reproduced some salient features of social networks, it lacks other realistic considerations such as the geographical dimension~\cite{Barthelemy2011Spatial} and multiplexity~\cite{Kivela2014Multilayer, Boccaletti2014Structure}. In order to consider such factors, Murase et al.~\cite{Murase2014Multilayer} extended the WSN model by assuming that each node can have connections in different layers for representing different types of connections. These nodes are also embedded into a two-dimensional geographical space, which enables to define the geographical distance between nodes as used in cases when the nodes choose other nodes that are geographically close to them for the GA mechanism. The locality of interaction naturally leads to the correlation between different layers and to a multilayer Granovetterian community structure. In the absence of interlayer correlations, the generated network does not have a community structure when merged into a single-layer network.

Since the ND mechanism, as the only mechanism eliminating links, could be too abrupt for representing the real dynamics of social networks, alternative mechanisms focusing on the links, namely, link deletion (LD)~\cite{Marsili2004Rise} and link aging, have been studied within the framework of the WSN model~\cite{Murase2015Modeling}. The LD mechanism is implemented by removing randomly chosen links irrespective of their weights, while for the link aging each link weight gradually decays and is removed when the link weight gets smaller than some threshold value. Different link termination mechanisms lead to different network structures, e.g., see Fig.~\ref{fig:snapshot}(top right) for the case of link aging. Three link termination mechanisms (ND, LD, and aging) are not exclusive to each other but may work simultaneously in reality. Note that the WSN model with the link deletion mechanism has been used to study the effect of the so-called channel selection on the network sampling~\cite{Torok2016What}.

Finally, Murase et al.~\cite{Murase2019Structural} extended the WSN model to incorporate the effect of homophily or the tendency of individuals to associate and bond with similar others~\cite{McPherson2001Birds}. For this, each node is assigned a set of features, similarly to the Axelrod model and its variants~\cite{Axelrod1997Dissemination, Centola2007Homophily, Vazquez2007Timescale, Vazquez2007Nonmonotonicity, Gandica2011Clustersize, Tilles2015Diffusion, Min2017Fragmentation}. Precisely, each node is represented by a feature vector that is a collection of traits for each feature. These features may relate to gender, ethnicity, language, religion, etc. The GA and LA mechanisms apply only to the nodes sharing the same trait for a randomly chosen feature per interaction. The number of features represents the social complexity of the population, while the number of traits per feature represents the heterogeneity of the population. Depending on the number of features and the number of traits per feature, the structural transition of the network from an overlapping community structure to a segregated, non-overlapping community structure is observed (see Fig.~\ref{fig:snapshot}(bottom)). 

\begin{algorithm*}[ht]
    \caption{\label{code:GWSN}
        Pseudocode for the GWSN model
    }
    \begin{algorithmic}[1]
    \For{$i=1,\ldots,N$} \Comment{Initialization}
        \State Initialize the node $i$'s location $(x_i,y_i)$ and its feature vector $(\sigma_i^1,\ldots,\sigma_i^F)$.
        \For{$j=1,\ldots,i-1,i+1,\ldots,N$}
            \State $w_{ij}\leftarrow 0$.
        \EndFor
    \EndFor
    
    \For{$t = 1, \ldots, t_{\rm max}$}
        \For{$i=1,\ldots,N$} \Comment{Global Attachment}
            \If{The node $i$ has no neighbor OR $\mathrm{rand}() < p_r $}
                \State Randomly choose a feature $f$ from $\{1,\ldots,F\}$.
                \State Obtain a set $\Gamma_i^f$ of nodes whose $f$th feature is the same as $\sigma_i^f$ and that are not connected to $i$.
                \If{$\Gamma_i^f$ is not empty}
                    \State Randomly choose a node $j$ from $\Gamma_i^f$ with the probability proportional to $r_{ij}^{-\alpha}$ (Eq.~\eqref{eq:geog}).
                    \State{Create a link between $i$ and $j$ and $w_{ij}\leftarrow w_0$.}
                \EndIf
            \EndIf
        \EndFor
        
        \For{$i=1,\ldots,N$} \Comment{Local Attachment}
            \State{Randomly choose a feature $f$ from $\{1,\ldots,F\}$.}
            \State{Obtain a set $\Lambda_i^f$ of $i$'s neighbors whose $f$th feature is the same as $\sigma_i^f$.}
            \If{$\Lambda_i^f$ is not empty} 
                \State{Randomly choose a node $j$ from $\Lambda_i^f$ with the probability proportional to $w_{ij}$ (Eq.~\eqref{eq:pLAij}).}
                \State{$w_{ij} \leftarrow w_{ij} + w_r$.}
                \State{Obtain a set $\Lambda_{j,\sim i}^f$ of $j$'s neighbors but $i$ whose $f$th feature is the same as $\sigma_j^f$.}
                \If{$\Lambda_{j,\sim i}^f$ is not empty} 
                    \State{Randomly choose a node $l$ from $\Lambda_{j,\sim i}^f$ with the probability proportional to $w_{jl}$ (Eq.~\eqref{eq:pLAjl}).}
                    \State{$w_{jl} \leftarrow w_{jl} + w_r$.}
                    \If{Nodes $i$ and $l$ are connected}
                        \State{$w_{il} \leftarrow w_{il} + w_r$.}
                    \ElsIf{$\mathrm{rand}() < p_{\Delta}$}
                        \State{Create a link between $i$ and $l$ and $w_{il}\leftarrow w_0$.}
                    \EndIf
                \EndIf
            \EndIf
        \EndFor
        
        \For{$i = 1, \ldots, N$} \Comment{Node Deletion}
            \If{$\mathrm{rand}() < p_{nd}$}
                \State{Remove all links adjacent to the node $i$ (i.e., $w_{ij}\leftarrow 0$ for every neighbor $j$).}
            \EndIf
        \EndFor
        \For{$ij$ in $\{ij|w_{ij}>0\}$}
            \If{$\mathrm{rand}() < p_{ld}$}  \Comment{Link Deletion}
                \State{Remove the link $ij$ (i.e., $w_{ij}\leftarrow 0$).}
            \Else
                \State{$w_{ij} \leftarrow w_{ij} \times (1-A)$} \Comment{Link Aging}
                \If{$w_{ij} < w_{th}$}
                    \State{Remove the link $ij$ (i.e., $w_{ij}\leftarrow 0$).}
                \EndIf
            \EndIf
        \EndFor
    \EndFor
    \end{algorithmic}
\end{algorithm*}

\begin{figure*}[!t]
\begin{center}
\includegraphics[width=0.9\textwidth]{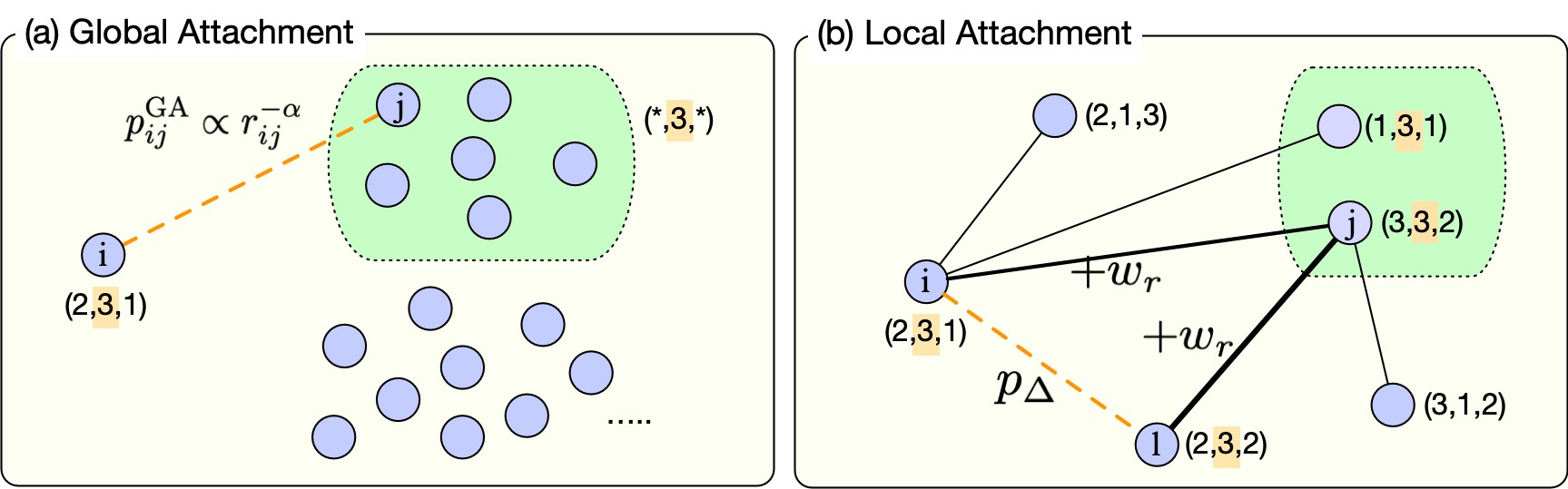}
\end{center}
\caption{Schematic diagrams of (a) the global attachment (GA) and (b) the local attachment (LA) of the GWSN model. Each node is described by a feature vector, e.g., $\vec\sigma_i=(2,3,1)$ for the node $i$. For each update by the GA and LA mechanisms, a feature is randomly chosen, e.g., $f=2$, then the node $i$ interacts only with other nodes, say $j$, with $\sigma_j^2=\sigma_i^2=3$. For the GA mechanism, the dependence on the geographical distance between nodes, $r_{ij}$, is controlled by the parameter $\alpha$ (see Eq.~\eqref{eq:geog}). For the LA mechanism, the triadic closure occurs with the probability $p_{\Delta}$ and the involved links increase their weights by $w_r$.}
\label{fig:schematic_diagram}
\end{figure*}

In addition, we remark that the temporal network version of the WSN model was studied by incorporating the timings of interactions by the GA and LA mechanisms~\cite{Jo2011Emergence}. This temporal WSN model generates not only the Granovetterian community structure but also bursty interaction patterns~\cite{Karsai2018Bursty}.

In this paper we treat the WSN model by considering all generalizations including the geographical dimension, various link termination mechanisms, and the homophilic interaction at the same time, which is called a generalized weighted social network (GWSN) model hereafter.

\subsection{Model definition}

We define a generalized weighted social network (GWSN) model for generating realistic social networks. Let us consider a network of $N$ nodes. Each individual node $i=1,\ldots,N$ is located in a fixed position $(x_i,y_i)$ that is randomly selected in a unit square $[0,1]\times[0,1]$ with periodic boundary conditions. The Euclidean distance between two nodes $i$ and $j$ is denoted by $r_{ij}$. The node $i$ is characterized by a feature vector of $F$ components, i.e., $\vec\sigma_i=(\sigma_i^1,\ldots,\sigma_i^F)$ with $\sigma_i^f\in\{1,\ldots,q\}$ for each $f\in \{1,\ldots,F\}$. The value of $\sigma_i^f$ is called a trait. The value of each $\sigma_i^f$ is uniformly randomly selected in the beginning of the simulation and then fixed throughout. In reality the different feature may have different numbers of traits, whereas we  assume for simplicity that all features have the same number of traits. By the homophilic interaction rule for a randomly chosen feature $f$, only a pair of nodes with the same trait, i.e., $\sigma_i^f=\sigma_j^f$, are allowed to interact with each other. A link weight between nodes $i$ and $j$ is denoted by $w_{ij}$, and $w_{ij}=0$ indicates the absence of the link between them.

The implementation of the GWSN model is summarized as a pseudocode in Algorithm~\ref{code:GWSN}. A simulation starts with an empty network of size $N$. For each Monte Carlo time step $t$, every node updates its neighborhood by sequentially applying the GA and LA mechanisms as well as three link termination mechanisms, i.e., ND, LD, and aging.

For the GA mechanism, let us consider a focal node $i$ and a randomly chosen feature $f$ from $\{1,\ldots,F\}$. We obtain a set of nodes whose $f$th feature has the same value as $\sigma_i^f$ and that are not connected with the node $i$, precisely,
\begin{align}
    \Gamma_i^f=\{j| \sigma_j^f=\sigma_i^f\ \&\ w_{ij}=0\}.
    \label{eq:definition_Si}
\end{align}
One of nodes in $\Gamma^f_i$, say $j$, is randomly chosen with the probability given by
\begin{align}
    p^{_{\rm GA}}_{ij} = \frac{r_{ij}^{-\alpha}}{\sum_{j' \in \Gamma_i^f}r_{ij'}^{-\alpha} }.
    \label{eq:geog}
\end{align}
Here the non-negative parameter $\alpha$ controls the locality of interaction. When $\alpha=0$, there is no geographical constraint. The larger the value of $\alpha$ is, the more likely nodes tend to make connections with geographically closer nodes. If the focal node $i$ has no neighbors, equivalently, if the node $i$'s degree is zero ($k_i=0$), it makes connection to the node $j$. Otherwise, if $k_i>0$, the link between nodes $i$ and $j$ is created with a probability $p_r$. The initial weight of the new link is set to $w_0$. See Fig.~\ref{fig:schematic_diagram}(a) for an example of the GA mechanism.

For the LA mechanism, if the focal node $i$ has neighbors, it randomly chooses one of its neighbors $j$ with $\sigma_j^f=\sigma_i^f$ with the probability proportional to the link weight $w_{ij}$, precisely,
\begin{align}
    p^{_{\rm LA}}_{ij}=\frac{w_{ij}}{\sum_{j'\in \Lambda_i^f}w_{ij'}}.
    \label{eq:pLAij}
\end{align}
Here $\Lambda_i^f$ denotes a set of $i$'s neighbors whose $f$th feature has the same value as $\sigma_i^f$. Then the node $j$ randomly chooses one of its neighbors but $i$, say $l$, with $\sigma_l^f=\sigma_j^f$, with the probability proportional to the link weigth $w_{jl}$.
\begin{align}
    p^{_{\rm LA}}_{jl}=\frac{w_{jl}}{\sum_{l'\in \Lambda_{j,\sim i}^f} w_{jl'}}.
    \label{eq:pLAjl}
\end{align}
Here $\Lambda_{j,\sim i}^f$ denotes the set of $j$'s neighbors, but $i$, whose $f$th feature has the same value as $\sigma_j^f$. If nodes $i$ and $l$ are not connected to each other, the link between them is created with the probability $p_{\Delta}$ and the initial weight of the new link is set to $w_0$. In addition, the weights of all involved links are increased by the reinforcement parameter $w_r$, irrespective of whether a new link is created or not. Figure~\ref{fig:schematic_diagram}(b) shows an example of the LA mechanism.

For the link termination, we sequentially apply all of ND, LD, and aging mechanisms. At each time step $t$, each node is replaced by an isolated node with the probability $p_{nd}$. In addition, each link in the network is deleted with the probability $p_{ld}$. For all links that are not deleted, the weight of each link is multiplied by a factor $1-A$ and the links whose weights are below a threshold value, denoted by $w_{th}$, are deleted. Here the parameter $A\in [0,1]$ controls the speed of aging.

We remark that previously studied modifications of the WSN model correspond to this GWSN model with some parameters set to certain values. For example, if each feature in the feature vector can have only one value, i.e., if $q=1$ for all $f\in\{1,\ldots,F\}$, every node can interact with every other node as $\sigma_i^f=1$ for all $i$ and $f$. It means that the GWSN model reduces to the model without the homophilic constraint. In addition, setting $p_{nd}$, $p_{ld}$, or $A$ to zero nullifies the ND, LD, or aging mechanism, respectively.

The input parameters of the GWSN model are summarized in Table~\ref{tab:input_parameters}, together with the sampling ranges of those parameters for the simulation. Here we respectively fix the values of $w_0$ and $w_{th}$ to $1$ and $0.5$, as they set scales only.

\begin{table}[!t]
  \caption{Notations and sampling ranges of the input parameters of the GWSN model. Parameters indicated with the asterisk ($^{\ast}$) take the values sampled uniformly on the logarithmic scale, while the others take the values sampled uniformly on the linear scale. Parameters indicated with the dagger ($^{\dagger}$) take only integer values. Values of $w_0$ and $w_{th}$ are fixed.
  \label{tab:input_parameters}}
  \begin{tabular}{|l|p{0.55\columnwidth}|l|}
  \hline
  Symbol & Input parameter & Sampling range \\
  \hline
  $N$ & Number of nodes & $[2000, 5000]^{\dagger}$ \\
  $F$ & Number of features & $[1, 10]^{\dagger}$ \\
  $q$ & Number of traits per feature & $[1, 10]^{\dagger}$ \\
  $\alpha$ & Dependence on the geographical distance & $[0, 4]$ \\
  $p_{r}$ & Probability of global attachment & $[10^{-5}, 10^{-2}]^{\ast}$ \\
  $p_{\Delta}$ & Probability of triadic closure & $[10^{-3}, 10^{-1}]^{\ast}$ \\
  $w_r$ & Link reinforcement in the local attachment & $[0, 2]$ \\
  $p_{nd}$ & Probability of node deletion & $[10^{-5}, 10^{-2}]^{\ast}$ \\
  $p_{ld}$ & Probability of link deletion & $[10^{-6}, 10^{-3}]^{\ast}$ \\
  $A$ & Speed of link aging & $[10^{-5}, 10^{-1}]^{\ast}$ \\

  \hline
  $w_0$ & Initial weight of new links & $1$ \\
  $w_{th}$ & Threshold weight for removing aged links & $0.5$\\
  
  \hline
  \end{tabular}
\end{table}

\begin{table}[!t]
  \caption{
  Notations of network properties considered in this work.
  \label{tab:outputs}
  }
  \begin{tabular}{|l|p{0.76\columnwidth}|}
  \hline
  Symbol & Network property \\
  \hline
  $\langle k \rangle$ & Average degree \\
  $\rho_k$ & Degree assortativity coefficient \\
  $\langle w \rangle$ & Average link weight \\
  $C$ & Average clustering coefficient \\
  $O$ & Average link overlap \\
  $\rho_{ck}$ & Pearson correlation coefficient between local clustering coefficient and degree \\
  $\rho_{ow}$ & Pearson correlation coefficient between link overlap and weight \\
  $(1 - f_c^a) \langle k \rangle$ & Percolation transition point rescaled by the average degree: The critical fraction $f_c^a$ of links are removed in ascending order of weights \\
  $(1 - f_c^d) \langle k \rangle$ & Percolation transition point rescaled by the average degree: The critical fraction $f_c^d$ of links are removed in descending order of weights \\
  \hline
  \end{tabular}
\end{table}

\subsection{Network properties}

For the systematic comparison between the generated networks by the GWSN model and the empirical networks, we adopt several quantities characterizing network properties as described below.
\begin{itemize}
    \item Average degree $\langle k\rangle$. It is defined as $\sum_{i=1}^N k_i/N$.
    \item Degree assortativity coefficient $\rho_k$. It is defined as $\rho_k=\sum_{kk'}kk'(e_{kk'}-q_kq_{k'})/\sigma^2_q$, where $e_{kk'}$ is the fraction of links connecting nodes with degrees $k$ and $k'$, $q_k$ is the distribution of the excess degree $k$, and $\sigma^2_q$ is the variance of $q_k$~\cite{Newman2002Assortative, Newman2003Mixinga}. It characterizes the preference for nodes to attach to others with similar degree.
    \item Average link weight $\langle w\rangle$. It is defined as $\sum_{ij} w_{ij}/L$, where $L$ is the number of links having positive weights.
    \item Average clustering coefficient $C$. It is defined as $\sum_{i=1}^N c_{i}/N$, where $c_i$ is the local clustering coefficient for the node $i$. $c_i$ is the number of links between $i$'s neighbors divided by $k_i(k_i-1)/2$~\cite{Watts1998Collective}.
    \item Average link overlap $O$. It is defined as $\sum_{i,j} o_{ij}/L$, where $o_{ij}$ is the link overlap for the link $ij$. $o_{ij}$ is defined as $\frac{n_{ij}}{(k_i-1)+ (k_j-1)-n_{ij}}$ with $n_{ij}$ being the number of neighbors common to both $i$ and $j$~\cite{Onnela2007Analysis, Onnela2007Structure}.
    \item Pearson correlation coefficient between local clustering coefficient and degree $\rho_{ck}$. This quantity measures correlation between $c_i$ and $k_i$ for nodes.
    \item Pearson correlation coefficient between link overlap and weight $\rho_{ow}$. This quantity measures correlation between $o_{ij}$ and $w_{ij}$ for links.
    \item Percolation transition points rescaled by the average degree. The Granovetterian community structure can be analyzed by means of the link percolation method~\cite{Onnela2007Analysis, Onnela2007Structure}. For this, links are sorted according to their weights. For the ascending percolation, the weakest links are removed one by one from the network to see how many links need to be removed to disintegrate the global structure of the network. $f_c^a$ denotes the fraction of links removed when the disintegration occurs. Thus, it is called ``transition point''. $f_c^d$ is similarly defined but by removing the strongest links one by one from the network. Significant difference between $f_c^a$ and $f_c^d$ implies the existence of the Granovetterian community structure. In our work we have recorded the fraction of remaining links at the percolation transition multiplied by the average degree for both ascending and descending link 
    percolation, denoted by $(1 - f_c^a) \langle k \rangle$, and $(1 - f_c^d) \langle k \rangle$, respectively. Generally speaking, the fraction of remaining links multiplied by the average degree indicates the average number of remaining neighbors per node.
\end{itemize}

\subsection{Remarks}

The GWSN model has a large number of input parameters as listed in Table~\ref{tab:input_parameters}, so that understanding the consequences of the model mechanisms and their effects on the statistical properties of the generated networks becomes a formidable task. So far each of the above mechanisms has been studied independently for simplicity and theoretical tractability~\cite{Murase2014Multilayer, Murase2015Modeling, Murase2019Structural}, however, these mechanisms might coexist and interfere in reality. The non-trivial interplay between such mechanisms makes the analysis of the model highly complex, hence the behavior of the GWSN model cannot be simply predicted based on the previous studies.

In order to overcome such difficulties we will use machine learning technique to perform regression analysis, as a new, general method to handle models with a large number of parameters. By using deep learning we will investigate the input-output relationships, where the inputs are model parameters (Table~\ref{tab:input_parameters}) and outputs are network properties (Table~\ref{tab:outputs}), respectively.

\section{Deep Learning Based Parameter Search}\label{sec:analysis}

\subsection{Regression analysis}

Since the GWSN model has a fairly large number of parameters, it is not easy to understand the model behavior when parameter values change simultaneously, which hinders tuning the parameter values to empirical data. In order to overcome this difficulty, we conduct a regression analysis using machine learning as follows. We first run the simulation of the model a large number of times for sampling the parameter space defined in Table~\ref{tab:input_parameters} uniformly randomly. We execute these simulations on the supercomputer Fugaku using four thousand nodes (192k CPU cores) for eight hours and used CARAVAN framework~\cite{murase2018caravan} for the job scheduling. Each simulation is implemented as an OpenMP~\footnote{\url{https://www.openmp.org/}} parallelized function running on four threads while the whole program runs as an MPI~\footnote{\url{https://www.mpi-forum.org/}} job. For each set of parameter values, we conduct five independent Monte Carlo runs with different random number seeds. For each generated network, we measure the network properties, such as the average degree and the average link weight, as listed in Table~\ref{tab:outputs}. The network quantities are measured every $5,000$ steps until the simulation finishes at $t = 50,000$, thus yielding $10$ entries for each run. Most simulations typically finish in $10$ minutes, however, we find very rare cases that are exceptionally computer time demanding. To avoid such edge cases, we abort the simulation if the average degree exceeds a threshold $k_{th} = 150$ (inspired by Ref.~\cite{Hill2003Social}) since the network is no longer regarded as a sparse network for the network sizes used (see Table~\ref{tab:input_parameters}). The number of features are $11$ since there are $10$ input parameters in addition to the simulation time step $t$. As a result, we obtained training data of about $25$ million entries ($0.5$ million parameter sets).

The sets of input parameters and output network properties are then used to train an artificial neural network (ANN) for the regression analysis. We use an ANN with three hidden layers each of which consists of $n_{\rm unit} (=30)$ fully connected rectifier linear unit (ReLU) with the dense output layer with the linear activation function. We chose ANN because it is one of the standard ways for meta-modeling and it indeed showed a significantly better performance for our case compared to the classical regression methods including Ridge, Lasso, and ElasticNet regression using up to third-order polynomial features. All input data are pre-processed by scaling each of the input ranges into the unit interval. The input parameters indicated by the asterisk ($^\ast$) in Table~\ref{tab:input_parameters} are scaled after taking the logarithm. ANNs are trained to minimize the mean squared error for $2,000$ epochs using the Adam optimization algorithm with a batch size of $200$. We obtained these hyper-parameters by the grid search over the number of hidden layers in $[2,3]$, the number of units in each layer in $[15,30]$, and the learning rate in $[0.001,0.01]$. The mean squared errors for the test set that are obtained independently from the training set are summarized in Table~\ref{tab:results}. The accuracy of the regression is good enough for our purpose although a better performance could be obtained by a more extensive tuning of hyper-parameters or other regression methods such as support vector regression, Gaussian process, or Kriging~\cite{Ghiasi2018comparative, Zhao2010comparative, Wang2007Review}. We used OACIS for the tuning of hyper-parameters~\cite{Murase2014Tool, Murase2017Opensource}. We conducted  regression for each output feature in Table~\ref{tab:outputs}. For the analysis of the link weights, we take the logarithm of $\langle w \rangle$ since its scale differs by orders of magnitude. Keras~\cite{chollet2015keras}, TensorFlow~\cite{tensorflow2015-whitepaper}, and scikit-learn~\cite{scikit-learn} were used for the implementation. The source code for both the simulation and the analysis is available online~\footnote{\url{https://github.com/yohm/GWSN_metamodeling}}.

\begin{table}[ht]
  \caption{
  Mean squared error (MSE) of the meta-models obtained for each output feature. The variances of the test dataset are also shown as a reference.
  \label{tab:results}
  }
  \begin{tabular}{|l|l|l|}
  \hline
  Network property & MSE & Variance \\
  \hline
  $\langle k \rangle$ & 3.22 & 676.4\\
  $\rho_k$ & 0.000607 & 0.0162 \\
  $\log_{10} \langle w \rangle$ & 0.000414 & 0.423 \\
  $C$ & 0.000363 & 0.0309 \\
  $O$ & 0.000247 & 0.0175 \\
  $\rho_{ck}$ & 0.00170 & 0.0730 \\
  $\rho_{ow}$ & 0.000647 & 0.0379 \\
  $(1 - f_c^a) \langle k \rangle$ & 0.122 & 1.23 \\
  $(1 - f_c^d) \langle k \rangle$ & 0.076 & 0.208 \\
  \hline
  \end{tabular}
\end{table}

\begin{figure*}[!t]
\begin{center}
\includegraphics[width=0.9\textwidth]{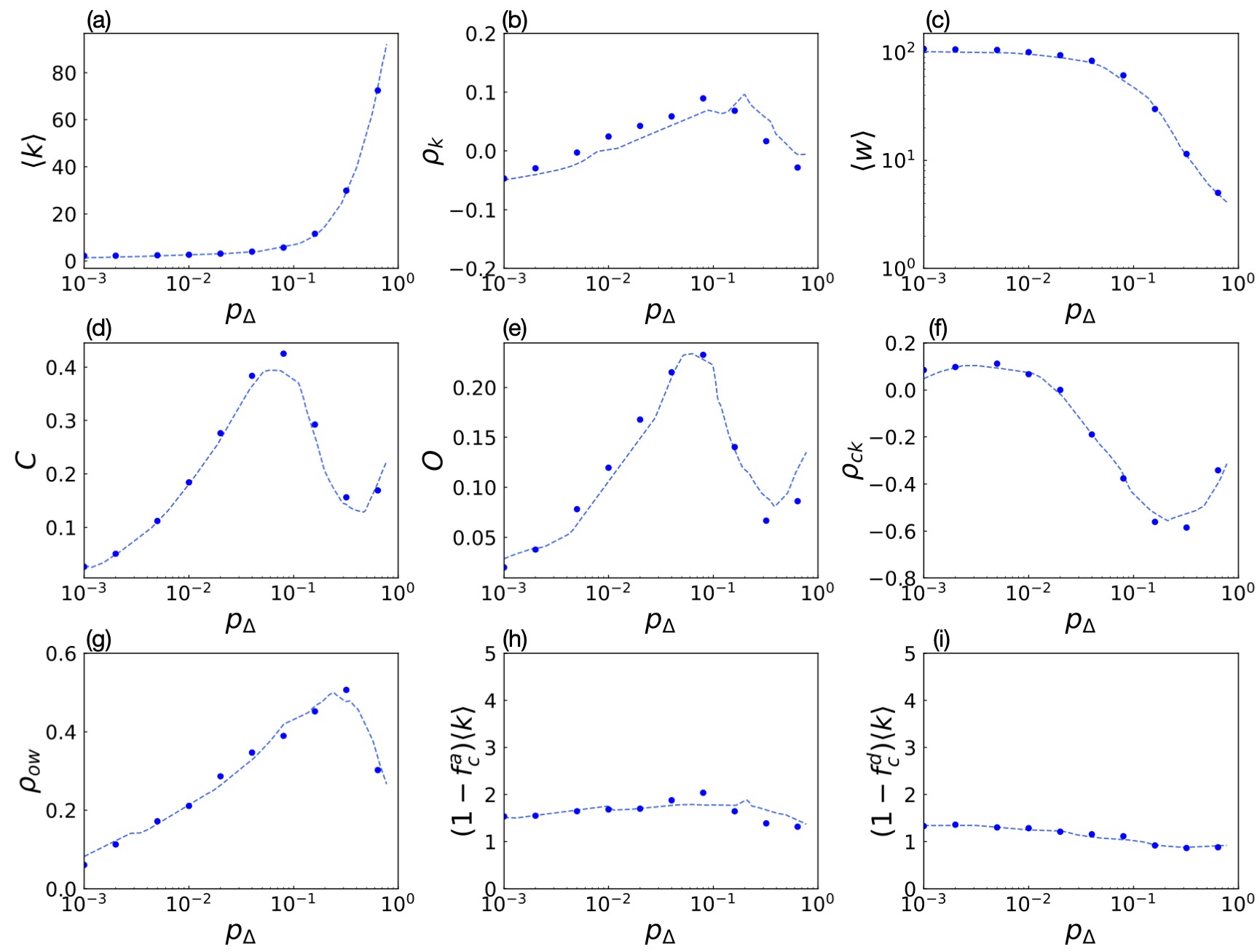}
\end{center}
\caption{
Comparison between the predictions by machine learning and the simulations. Each panel shows how each network property of the generated networks changes with respect to the value of the parameter for the triadic closure, $p_{\Delta}$. In each panel the dashed curve shows the predictions by machine learning while the filled circles indicate the simulation results averaged over five independent runs, showing that the machine learning predicts the model outputs quite well. Other parameter values are fixed at $N=2,000$, $F = 2$, $q = 2$, $\alpha = 0$, $p_r = 0.001$, $w_r = 1$, $p_{nd} = 0.001$, $p_{ld} = 0.01$, $A = 0.003$, and $t = 50,000$.
}
\label{fig:regression}
\end{figure*}

An example of the results of the regression analysis is shown in Fig.~\ref{fig:regression}. In Fig.~\ref{fig:regression}(a), the average degree $\langle k\rangle$ predicted by the machine learning as a function of $p_{\Delta}$ is shown as the dashed curve when $N=2,000$, $F = 2$, $q = 2$, $\alpha = 0$, $p_r = 0.001$, $w_r = 1$, $p_{nd} = 0.001$, $p_{ld} = 0.01$, $A = 0.003$, and $t = 50,000$ are used. In the same figure, the simulation results are depicted by filled circles. Note that the prediction by the machine learning is not just an interpolation of these simulation results since these simulation results were not included in the training data but conducted separately. Figure~\ref{fig:regression} also includes the results for other network properties such as $\rho_k$, $\langle w \rangle$, $C$, $O$, $\rho_{ck}$, $\rho_{ow}$, $(1-f_c^a)\langle k \rangle$, and $(1-f_c^d)\langle k \rangle$. While the behavior of the average degree $\langle k\rangle$ as a function of $p_\Delta$ is easy to interpret, the other quantities show complex non-monotonic dependence on $p_{\Delta}$. For instance, as shown in Fig.~\ref{fig:regression}(d), the average clustering coefficient $C$ increases with $p_{\Delta}$ for the ranges of $p_{\Delta} < 0.1$ and $p_{\Delta} > 0.3$, while it decreases with $p_{\Delta}$ for the range of $0.1 < p_{\Delta} < 0.3$. Other non-monotonic dependencies on $p_{\Delta}$ are also observed for other network properties. As this example demonstrates, the meta-models reproduce the model outputs, which is useful not only for quantitative parameter tuning but also for understanding the properties of the model. We provide an online interactive chart that allows users to observe the responses of the network properties to the input parameters~\footnote{\url{https://yohm.github.io/GWSN_metamodeling}}.

\subsection{Sensitivity analysis}

\begin{figure*}[!t]
\begin{center}
\includegraphics[width=0.9\textwidth]{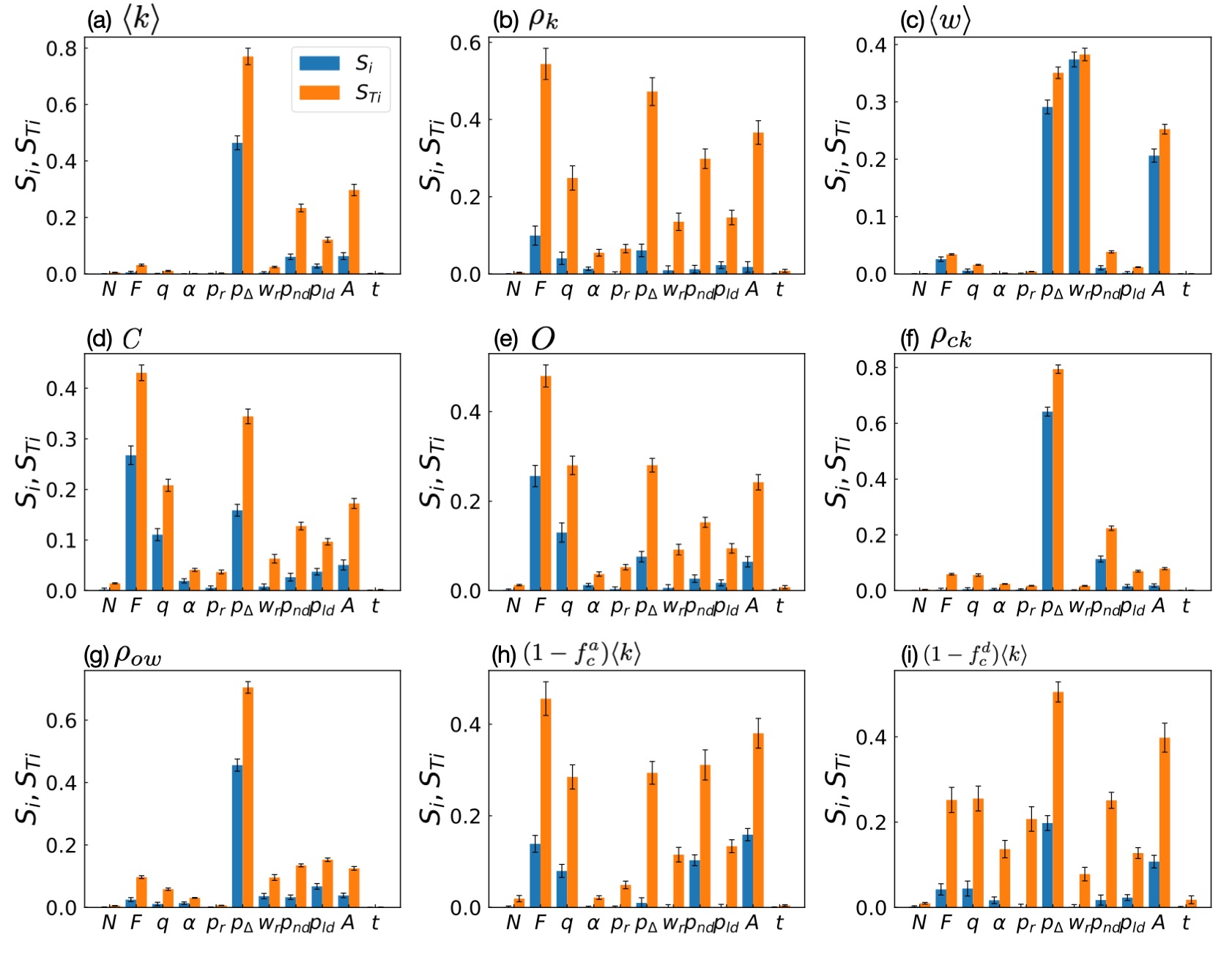}
\end{center}
\caption{
Sensitivity analysis of the meta-model. The first-order index $S_i$ and the total-effect index $S_{Ti}$ of each input parameter (Table~\ref{tab:input_parameters}) against each output (Table~\ref{tab:outputs}) are shown with error bars indicating the $95\%$ confidence intervals. See Eqs.~\eqref{eq:Si} and~\eqref{eq:STi} for definitions of $S_i$ and $S_{Ti}$, respectively.
}
\label{fig:sensitivity}
\end{figure*}

The input parameters of the model are not equally important. While some of them have a major impact on the output, others do not alter the results significantly. Sensitivity analysis is a method to evaluate how the uncertainty in the output of a model or system can be divided and attributed to the uncertainty in its inputs, namely, it tells which input parameters are important and which are not. While several methods have been proposed for sensitivity analysis~\cite{Saltelli2007Global}, here we adopt the variance-based sensitivity analysis proposed by Saltelli et al.~\cite{Saltelli2010Variance, Sobol'2001Global}. As we have seen in Fig.~\ref{fig:regression}, the GWSN model shows highly non-linear dependency on the input parameters. Furthermore, simultaneous perturbations of two or more input parameters often cause greater variations in the output than the sum of the variations caused by each of the perturbations alone. The variance-based method is effective for such models because it measures sensitivity across the entire input space (i.e., it is a global method), it deals with nonlinear responses, and it takes into account the effect of interactions in non-additive systems. While the variance-based methods generally require a larger number of sampling than other methods, they are easily calculated once we developed a meta-model as is done in the previous section.

In variance-based sensitivity analysis, the sensitivity of the output to an input variable is quantified by the amount of variance caused by the fluctuation of the input. Consider a generic model $Y = f(X_1, X_2, \ldots, X_n)$, i.e., the output $Y$ is a function of $n$ input variables $\mathbf{X}$. Due to the uncertainties of the input parameters, each of which considered independent random variables following the probability distributions $p_i(X_i)$, the output $Y$ has also some uncertainty. Let us denote the variance of $Y$ as $V[Y] = E[Y^2] - E[Y]^2$, where the operator $E[\cdot] \equiv \int \cdot p(\mathbf{X}) d\mathbf{X}$ indicates the expected value. We calculate the two indices representing the sensitivity of $Y$ to $X_i$, namely, the first-order sensitivity index $S_{i}$ and the total-effect sensitivity index $S_{Ti}$. These indices are defined as
\begin{align}
\label{eq:Si}
    S_{i}     &= \frac{V_i\left[ E_{\sim i}\left[Y\right] \right]}{V(Y)}, \\
    \label{eq:STi}
    S_{Ti} &= \frac{E_{\sim i}\left[ V_{i}\left[Y\right] \right]}{V(Y)},
\end{align}
where $E_i[\cdot]$ and $E_{\sim i}[\cdot]$ denote the expected values averaged over $X_i$ and over all input variables but $X_i$, respectively. The operator $V_i[\cdot] \equiv E_i[\cdot ^2] - E_i[\cdot]^2$ denotes the variance taken over $X_i$. The first-order index $S_{i}$ indicates the expected reduction of variance when the input $X_i$ could be fixed. The interactions between the other inputs are not taken into account. On the other hand, the total-effect index $S_{Ti}$ tells us the importance of $i$th input taking into account all the higher-order interactions in addition to the first-order effect. We used SAlib python package to calculate the sensitivity indices~\cite{Herman2017}.

Figure~\ref{fig:sensitivity} shows the results of the sensitivity analysis. In Fig.~\ref{fig:sensitivity}(a) we show the first-order and the total-effect indices for the average degree $\langle k \rangle$. The figure indicates that the average degree is the most sensitive against the change in $p_{\Delta}$ whether $p_{\Delta}$ changes independently or with other parameters at the same time. The next important factors for $\langle k \rangle$ are the parameters for link termination, i.e., $p_{nd}$, $p_{ld}$, and $A$. Since they show large values of $S_{Ti}$, these parameters seem to strongly interact with other parameters. The average link weight $\langle w \rangle$ displays a different tendency as shown in Fig.~\ref{fig:sensitivity}(c). It shows strong dependency mainly on $w_r$ and $A$ in addition to $p_{\Delta}$, which is reasonable since $w_r$ and $A$ control the link reinforcement and the link aging, respectively.

If we look at the indices for clustering coefficient shown in Fig.~\ref{fig:sensitivity}(d), we find a different behavior. The parameters for the homophilic interaction, i.e., $F$ and $q$, have high values of $S_i$ and $S_{Ti}$, indicating that these parameters affect the clustering properties of the emergent networks, while they play minor role in determining the average degree as shown in Fig.~\ref{fig:sensitivity}(a). A similar tendency is observed for the average link overlap $O$ in Fig.~\ref{fig:sensitivity}(e); this is reasonable since both the clustering coefficient and the link overlap essentially quantify the frequency of closed triads.

Finally, some input parameters such as $N$, $t$, $\alpha$, and $p_r$ lead to relatively small values of $S_{Ti}$ for most outputs, indicating that these input parameters rarely affect the outputs even when they change with other parameters. The insignificance of $N$ and $t$ indicates that the finite-size effect is negligibly small and the simulations reach statistically stationary state for most cases. Consequently, the model can be simplified by fixing these parameters to arbitrary values within their sampling ranges.

\section{Summary and Discussion}\label{sec:summary}

In this paper we have focused on studying the formation of social network using an agent-based GWSN model, which incorporates several realistic extensions, such as homophilic interactions and link aging, to the original WSN model~\cite{Kumpula2007Emergence}. The effects of these additional mechanisms were studied independently in our previous papers~\cite{Murase2014Multilayer,Murase2015Modeling,Murase2019Structural} for simplicity and analytical tractability. However, in real world social networks these mechanisms coexist, which, if incorporated within one model, poses a challenge, because they often interact in non-trivial ways. In the framework of modelling this means that with the number of model parameters the number of possible combination effects increases super-linearly. Such difficulties are common for agent-based models~\cite{Bonabeau2002Agentbased, Sayama2015Introduction}. Thus, the use of agent-based models has been limited for long time to either research for qualitative understanding using relatively simplistic models, or research that uses complex models but does not fully explore the parameter space.

Instead of the traditional approach to build first a simple model and add elements one by one to study in sequence their consequences, we started from a generalized model with all the elements incorporated and then studied its behavior using large-scale computation and meta-modelling. Hence a massive number of simulations were performed using a supercomputer to sample a high-dimensional input space, and the outputs of the model were then used as training data for machine learning. We demonstrated that the meta-model, obtained by a deep multilayer perceptron, accurately reproduces the behavior of the model over a wide range of input parameters. Although massive computation is needed for the preparation of the training data, these simulations were efficiently executed using a supercomputer since they can be calculated independently and in parallel. Once the learning phase is completed, it is computationally quite cheap to calculate a prediction, which is useful for various purposes including the understanding of model behavior, tuning of parameter values, and sensitivity analysis.

As a demonstration, we conducted sensitivity analysis of the  meta-model and identified which input parameters are most influential on the output, i.e., on the properties of the generated networks. In addition, the sensitivity analysis tells us which parameters have negligible effect on the output, indicating that these parameters can be fixed to certain values in their sampling ranges without the loss of generality and for simplifying the model. Therefore, our computational approach enables us to study a model with a larger number of parameters in a systematic and quantitative way. Taking the full GWSN model with the traditional approach of parameter fitting would cause severe, if not unsolvable, difficulties, while our meta-modeling approach allowed us to study it with reasonable computational effort. 

There are a couple of open issues for future research. First of all, the GWSN model should be extended further. While we stepped forward from the simplest plausible model to more realistic ones, there is still a long way to go, considering the complexity of real social system. A further possible future direction is the extension of the output network characteristics. We quantified the network properties only by scalar values such as the average degree, yet we did not pay attention to more informative quantities, such as the degree distribution. This is mainly for the simplicity in developing meta-models, however the heterogeneity of the network, often characterized by the broad distributions, is of crucial importance in investigating not only the network properties but also dynamical processes taking place on those networks such as epidemic spreading~\cite{Pastor-Satorras2015Epidemic} and random walks~\cite{Masuda2017Random}. It is expected that meta-models approximating the distributions obtained from the agent-based models will be developed in the near future. Another promising future direction would be the extension of models toward temporal, multiplex, and higher-order networks~\cite{Holme2012Temporal, Kivela2014Multilayer, Battiston2020Networks} as there has been increasing demand to representing social Big Data in these terms. It would be straightforward to extend the meta-modeling approach of this paper beyond the simple pairwise interactions.

\begin{acknowledgments}
Y.M. acknowledges support from Japan Society for the Promotion of Science (JSPS) (JSPS KAKENHI; Grant no. 18H03621 and Grant no. 21K03362). H.-H.J. was supported by Basic Science Research Program through the National Research Foundation of Korea (NRF) funded by the Ministry of Education (NRF-2018R1D1A1A09081919). J.T. was supported by NKFIH (Grant Nos. OTKA K129124). J.K. was partially supported by EU H2020 Humane AI-net (Grant \#952026) and EU H2020 SoBigData++ (Grant \#871042). K.K. acknowledges support from EU H2020 SoBigData++ project and Nordforsk's Nordic Programme for Interdisciplinary Research project ``The Network Dynamics of Ethnic Integration". Y.M., H.-H.J., J.T., and J.K. are thankful for the hospitality of Aalto University. This research used computational resources of the supercomputer Fugaku provided by the RIKEN Center for Computational Science.
\end{acknowledgments}

%

\end{document}